\documentclass[10pt,conference]{IEEEtran}
\ifCLASSOPTIONcompsoc
  \usepackage[nocompress]{cite}
\else
  \usepackage{cite}
\fi
\ifCLASSINFOpdf
\else
\fi
\usepackage{authblk}
\usepackage{mathtools}
\DeclarePairedDelimiter\floor{\lfloor}{\rfloor}
\DeclarePairedDelimiter{\ceil}{\lceil}{\rceil}
\usepackage{amsmath,amssymb,amsfonts}
\usepackage{algorithm}
\usepackage{algpseudocode}
\usepackage{mdframed}
\usepackage{graphicx}
\usepackage{subfig}
\usepackage{multicol}
\usepackage{textcomp}
\usepackage{xcolor}

\def\BibTeX{{\rm B\kern-.05em{\sc i\kern-.025em b}\kern-.08em
    T\kern-.1667em\lower.7ex\hbox{E}\kern-.125emX}}
\newcommand{\comment}[1]{}

\newtheorem{theorem}{Theorem}
\newtheorem{lemma}{Lemma}
\newtheorem{problem}{Problem}

\newtheorem{corollary}{Corollary}

\usepackage{mdframed}

\begin{document}
%
\title{Minimal Overhead ARQ Sharing Strategies for URLLC in Multi-Hop Networks}

%
%
%
%

\author{Jaya Goel$^{*}$ and J. Harshan$^{\dagger, *}$\\
$^{*}$Bharti School of Telecom Technology and Management, Indian Institute of Technology Delhi, India.\\
$^{\dagger}$Department of Electrical Engineering, Indian Institute of Technology Delhi, India.
}


%
%

\markboth{Journal of \LaTeX\ Class Files,~Vol.~XX, No.~X, XXXX}%
{Shell \MakeLowercase{\textit{et al.}}: Bare Demo of IEEEtran.cls for Computer Society Journals}
%



\IEEEtitleabstractindextext{%
\begin{abstract}
The problem of achieving ultra-reliable and low-latency communication (URLLC) in multi-terminal networks has gained traction in the recent past owing to new wireless applications in vehicular networks. In the context of multi-hop networks, which is a classic example for multi-party communication, recent studies have shown that automatic-repeat-request (ARQ) based decode-and-forward (DF) strategies are suitable for URLLC since the idea of distributing a given number of ARQs across the nodes provides fine control on the features of reliability and latency. Inspired by these developments, in this work, we propose a cooperative ARQ sharing strategy for URLLC in multi-hop networks. At the heart of the proposed scheme lies the idea that every node is given the knowledge of the number of ARQs allotted to its preceding node in addition to the ARQs allotted to itself. As a result, each node only needs to count the number of unsuccessful attempts of its preceding node, and then borrow the unused ARQs, thereby improving the reliability feature with no compromise in the latency constraint. Using packet-drop-probability (PDP) as the reliability metric for the proposed cooperative strategy, we formulate an optimization problem of minimizing the PDP subject to a sum constraint on the total number of ARQs allotted across all the nodes. Supported by theoretical analysis on the behaviour of PDP, we present low-complexity algorithms to compute near-optimal ARQ distributions for our strategy, and show that our strategy outperforms the existing non-cooperative strategies.
\end{abstract} 

}

\maketitle

\IEEEdisplaynontitleabstractindextext

%
\IEEEpeerreviewmaketitle

\section{Introduction}
\label{sec:intro}

Signal design for ultra-reliable and low-latency communication (URLLC) has received much attention in the recent past for applications involving wireless devices that need to communicate its messages to an intended destination within a given deadline \cite{Pocovi2018, Ji2018}. Example applications include autonomous vehicle-to-vehicle/infrastructure (V2X) communication, wherein stringent deadlines are imposed on the round-trip delay between the vehicles and the infrastructure, beyond which either the messages are rendered stale or the deadline violation may lead to catastrophic consequences \cite{Schulz2017}. While the challenges of URLLC has been the topic of interest to communication over point-to-point channels, high reliability and low-latency constraints have also been studied in advanced multi-terminal settings such as multi-hop networks \cite{Changyang_She, Hong_Ren, Chen2018}. One such well known application of URLLC in multi-hop networks involves a network of Unmanned Aerial vehicles (UAVs), wherein power-limited UAVs operate as relays in coordinating mobility of autonomous vehicles. Among the many known protocols for multi-hop networks \cite{laneman_relay}, it has been recently shown that ARQ based DF strategies \cite{Wiemann2005, our_work_1} are well suited for URLLC in multi-hop networks as they provide fine control on the features of reliability as well as latency. Specifically, the number of ARQs allotted to a relay node provides control on the reliability feature (to combat the degrading effects of fading channels), whereas imposing a sum constraint on the total number of ARQs across the nodes provides control on the latency feature. Formally, let the processing time at each hop be denoted by $\tau_{p}$ seconds (which includes encoding and decoding time for the packet), and let the delay incurred due to re-transmission of packets at each link be denoted by $\tau_{d}$ seconds. Given that the wireless channels between successive relays are stochastic in nature, total number of packet re-transmissions before the packet reaches the destination is a random variable, denoted by $n$, and as a result, the end-to-end delay between the source and the destination is given by $n \times (\tau_{p} + \tau_{d})$ seconds. When the packet size and the decoding protocol are fixed, $\tau_{p}$ and $\tau_{d}$ are fixed, and therefore, when strict deadlines on end-to-end delay (denoted by $\tau_{total})$ are known, we can impose an upper bound on $n$, given by $q_{sum} = \floor{\frac{\tau_{total}}{\tau_{p}+\tau_{d}}}$, and then attempt to maximise the reliability feature by appropriately distributing the $q_{sum}$ ARQs across the nodes in the multi-hop network.

In the ARQ based DF strategy, although the source node and the intermediate relays are allotted a certain number of ARQs to facilitate successful transmission of the packet to the next node, imposing a sum constraint on the total number of ARQs results in a non-zero probability with which the packet does not reach the destination. Henceforth, throughout the paper, we refer to this probability as the packet-drop-probability (PDP). Since the multi-hop network is composed of multiple wireless links, the PDP is function of (i) the ARQs allotted to each link, (ii) the Line-of-Sight (LOS) component at each link, (iii) the underlying signal-to-noise-ratio (SNR), and importantly (iv) the underlying protocol of the network to achieve high reliability. Towards handling this problem statement, \cite{our_work_1} recently proposed a framework of non-cooperative strategy wherein each node uses the ARQs allotted to itself beyond which the packet is said to be dropped in the network. With such a model, \cite{our_work_1} addressed the problem of optimal allocation of ARQs across the nodes such that the PDP is minimized subject to a given sum constraint on the ARQs. Although the framework of non-cooperative strategy attempts to reduce the PDP by imposing latency-constraints in the form of an upper bound on the total number of ARQs, we observe that the non-cooperative strategy has a fundamental limit with which the PDP can be minimized. Motivated by this observation, in this paper, we explore whether an ARQ based DF strategy can be proposed with cooperation among the relay nodes so as to further increase the reliability when compared to the non-cooperative strategy with no relaxations on the latency constraints on the packets. Towards that direction, we make the following contributions in this paper:

\noindent 1) Under the class of ARQ based DF strategies for multi-hop networks, we propose a cooperative ARQ model, referred to as the semi-cumulative ARQ based DF strategy, wherein each transmitter in the network has the knowledge of the number of ARQs allotted to its preceding node in addition to the number of ARQs allotted to itself. We show that this cooperative framework assists a relay node in borrowing unused ARQs from the preceding node thereby increasing the reliability of the packets with no compromise in the latency constraints. We highlight that the benefits offered by the proposed cooperative strategy does not accompany additional overheads since every node only needs to count the number of failed attempts when decoding the packet received from the preceding node (see Section \ref{sec:network_model}).\\ 
2) For the proposed semi-cumulative ARQ based DF strategy, we address the problem of computing the optimal ARQ distribution that minimizes the PDP subject to a sum constraint on the total number of ARQs. Towards that direction, first, we use the Fibonacci series to derive closed-form expressions on the PDP of the semi-cumulative strategy for arbitrary $N$ and $q_{sum}$, and then formally prove that the proposed strategy outperforms the non-cooperative strategy in \cite{our_work_1}. We highlight that the task of deriving the PDP expression is a non-trivial contribution owing to the memory property introduced by the idea of borrowing unused ARQs of the preceding nodes (see Section \ref{sec:analysis}).\\  
3) To solve the PDP minimization problem, first, we prove that the problem of computing the optimal ARQ distribution for an $N$-hop network can be reduced to the problem of computing the optimal ARQ distribution for an $(N-2)$-hop network, thereby showcasing a substantial reduction in the complexity (see Section \ref{sec:opt_distribution}). Subsequently, generalizing the reduction approach, we propose two classes of low-complexity algorithms that can be used to compute near-optimal ARQ distributions for any $N$-hop network (see Section \ref{sec:proposed_method}). We also present extensive simulation results to showcase the efficacy of the proposed algorithms in terms of PDP reduction as well as computational complexity (see Section \ref{sec:Sims}).

\section{Semi-Cumulative Multi-Hop Network}
\label{sec:network_model}
\begin{figure}[h!]
\centering \includegraphics[scale = 0.27]{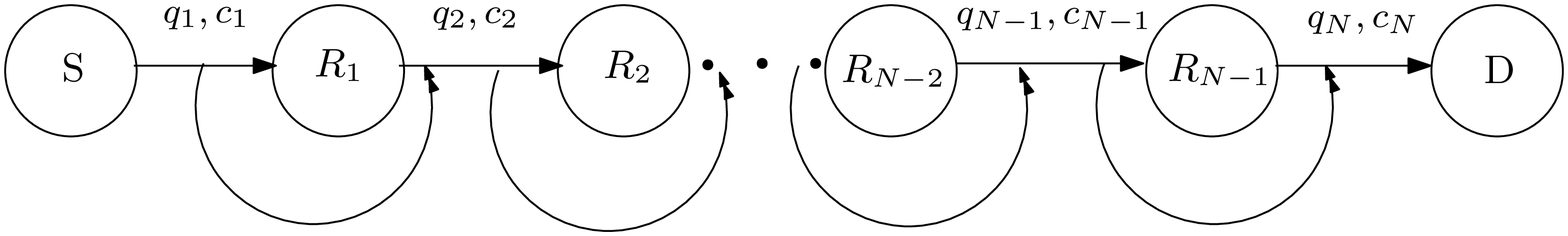}\centering{\caption{Illustration of an $N$-hop semi-cumulative scheme with channels dominated by LOS components.
\label{Networkmodel}}}
\end{figure}
Consider an $N$-hop network, as shown in Fig. \ref{Networkmodel}, wherein a source node intends to communicate its messages to a destination through a set of $N-1$ relay nodes that operate using an ARQ based DF strategy. In this model, the multi-hop network is characterized by the LOS vector $\mathbf{c}= \{c_{1}, c_{2},\ldots,c_{N}\}$ and the ARQ distribution $\mathbf{q}= \{q_{1}, q_{2},\dots,q_{N} \}$, such that $c_{i} \in [0,1]$ represents the LOS component of the fading channel of the $i$-th hop and $q_{i}$ represents the number of re-transmissions allotted to the transmitter of the $i$-th hop, for $1 \leq i \leq N$.

Formally, let $\mathcal{S} \subset \mathbb{C}^{K}$ denote the channel code employed at the source node of rate $R$ bits per channel use, i.e., $R = \frac{1}{K}\log_{2}(|\mathcal{S}|)$. Let $\mathbf{x} \in \mathcal{S}$ denote the packet (traditionally referred to as a codeword) transmitted over the multi-hop network such that $\frac{1}{K} \mathbb{E}[|\mathbf{x}|^{2}] = 1$. When $\mathbf{x}$ is transmitted over the $i$-th link, for $1 \leq i \leq N$, the corresponding received signal after $K$ channel uses is given by $\mathbf{y}_{i} = h_{i}\mathbf{x} + \mathbf{n}_{i} \in \mathbb{C}^{K}$, where $h_{i}$ is a quasi-static Ricean fading channel given by $h_{i} = \sqrt{\frac{c_{i}}{2}}(1+\iota)+\sqrt{\frac{(1-c_{i})}{2}}g_{i},$ such that $\iota = \sqrt{-1}$, $g_{i}$ is distributed as $\mathcal{CN}(0,1)$, $\mathbf{n}_{i}$ is the additive white Gaussian noise (AWGN) vector at the receiver of the $i$-th link, distributed $\mathcal{CN}(0,\sigma^{2}\mathbf{I}_{K})$. We assume that the receiver of each link has  perfect knowledge of its channel, however, there is no knowledge about the channel at the transmitter side. Furthermore, it is possible that the instantaneous mutual information of the channel may not support the transmission rate $R$ as the channel realization $h_{i}$ is random and does not change over $K$ channel uses. Therefore, if the mutual information of the channel is less than the transmission rate, then the receiver will not able to decode the packet correctly, and this event is referred to as the outage event whose probability is given by 
\begin{eqnarray}
\label{eq:outage_prob_link}
P_{i} = \mbox{Prob} \Big( R > \log_{2}(1+ |h_{i}|^{2}\alpha ) \Big) = \mbox{F}_{i}\left(\frac{2^{R}-1}{\alpha}\right),
\end{eqnarray}
where $\alpha =  \frac{1}{\sigma^{2}}$ is the average signal-to-noise-ratio (SNR) of the $i$-th link, $\mbox{F}_{i}(x)$ is the cumulative distribution function of $|h_{i}|^{2}$, defined as $\mbox{F}_{i}\left(\frac{2^{R}-1}{\alpha}\right) = 1-Q_{1}\Bigg(\sqrt{\frac{2c_{i}}{(1-c_{i})}}, \sqrt{\frac{2(2^{R}-1)}{\alpha(1-c_{i})}}\Bigg),$ such that $Q_{1}(\cdot,\cdot) $ is the first-order Marcum-Q function. Owing to the above mentioned outage events, a transmitter of the $i$-th hop is allotted $q_{i}$ number of re-transmissions in order to successfully forward the packet to the next node in the network. Despite using this ARQ based DF strategy, if a transmitter is unable to transmit the packet within $q_{i}$ number of ARQs, then the packet is said to be dropped in the network. Since the packet can be dropped at any hop of the network, we use PDP as the reliability metric of interest, which is defined as the fraction of packets that do not reach the destination. 

To achieve higher reliability than the ARQ based DF protocol in \cite{our_work_1}, we propose the semi-cumulative model, as shown in Fig. \ref{Networkmodel}, wherein every intermediate relay node can use residual ARQs unused by its previous node in the chain. This simple idea stems from the fact that although $q_{i}$ re-transmissions are allotted to a transmitter, the actual number of re-transmissions can be less than $q_{i}$ owing to the stochastic nature of the wireless channel. To facilitate this, we assume that every node has the knowledge of the number of ARQs given to its preceding node in addition to the ARQs allotted to itself. Since a given node can use unused ARQs from its previous node, the total number of ARQs used by it can be more than the number of ARQs allotted to it. As a result, the next node in the chain, despite knowing the number of ARQs allotted to its preceding node, does not know how long to wait for the successful transmission of the packet. To fix this, each intermediate node will have to wait for a fixed amount of time to receive the packet from its previous node beyond which the packet is said to be dropped in the network. Unlike the non-cooperative strategy of \cite{our_work_1}, in this method, an intermediate node can get more re-transmissions than the number of allotted to it just by listening to the number of failed attempts of the preceding node. Although each relay node is allowed  multiple transmissions (including the number of unused ARQs of its preceding node) to communicate the packet to the next node, there is a non-zero probability with the packet is dropped in the network since the sum of the ARQs allotted to all the nodes in the network is bounded, i.e., $\sum_{i=1}^{N}q_{i} = q_{sum}$. Henceforth, we denote the PDP of the semi-cumulative ARQ based DF strategy by $pdp_{sc, N}$, where $sc$ in the subscript highlights the semi-cumulative scheme, and $N$ denotes the number of hops in the network. Thus, in order to provide reliability along with low-latency constraint on the packets, in this paper, we propose to solve Problem \ref{opt_problem}, as shown below.

\vspace{0.5cm}
\begin{mdframed}
	\begin{problem}
		\label{opt_problem}
		For an $N$-hop network with a given LOS vector $\mathbf{c}$, a given SNR $\alpha = \frac{1}{\sigma^{2}}$, and a given $q_{sum}$, solve
		\begin{equation*}
		q_{1}^{*},q_{2}^{*},\ldots q_{N}^{*}=\arg\underset{q_{1},q_{2},\ldots q_{N}}{\text{min}}\ pdp_{sc, N}
		\end{equation*}
		$\text{subject to} ~q_{i}\in \mathbb{Z}_{+} \ \forall i, q_{i} \geq 0 \ \mbox{for~} i \geq 2, q_{i+1} \neq 0 \ \text{if} \ q_{i} = 0 \ \text{or}\ 1, \mbox{~and ~}\sum_{i = 1}^{N} q_{i} = q_{sum}, \mbox{~where ~} i \in \{1,2,\ldots N\}.$
	\end{problem}
\end{mdframed}
\vspace{0.5cm}

Towards solving Problem \ref{opt_problem}, in the next section, we derive an expression for the PDP of the semi-cumulative scheme, and then formally prove that the semi-cumulative scheme outperforms the non-cooperative ARQ strategy in \cite{our_work_1}.


\section{PDP Expression of Semi-Cumulative Scheme} 
 \label{sec:analysis}
 
 \begin{theorem}
 	\label{thm1}
 	The PDP expression for an $N$-hop semi-cumulative scheme is given by
 	\begin{equation}
 	pdp_{sc,N} = P_{1}^{q_{1}} + (1-P_{1})P_{2}^{q_{2}}F_{2} + \ldots + \prod_{i = 1}^{N-1}(1-P_{i})P_{N}^{q_{N}}F_{N},
 	\end{equation}
 	where $F_{j}$, for $2 \leq j \leq N$, is a function of $P_{1}, P_{2}, \ldots, P_{j-1}$ (as given in \eqref{eq:outage_prob_link}) and $q_{1}, q_{2}, \ldots, q_{j-1}$ that can be computed using Fibonacci series. 
 \end{theorem}
 \begin{IEEEproof}
 	For a $2$-hop network, the PDP expression, denoted by $pdp_{sc,2}$, can be written as $pdp_{sc,2} = pdp_{sc,1h} + pdp_{sc,2h}$, where $pdp_{sc,1h} = P_{1}^{q_{1}}$ and 
 	$pdp_{sc,2h} = (1-P_{1})P_{2}^{q_{2}}\bigg(\sum_{i=1}^{q_{1}} P_{1}^{q_{1}-i}P_{2}^{i-1}\bigg)$ represent the probability that the packet is dropped at the $j$-th hop for $j \in \{1,2\}$. From the definition of the semi-cumulative scheme, the first node does not have any preceding node to borrow ARQs whereas the second node can borrow unused ARQs from the first node. In the expression for $pdp_{sc,2h}$, the term $\sum_{i=1}^{q_{1}}P_{1}^{q_{1}-i}P_{2}^{i-1}$, henceforth referred to as $\beta_{E}^{(1,2)}$, captures the probability that the packet is dropped in the second link despite using the residual ARQs from the first link. These types of terms are known as external borrowing ARQ terms. In short, we can rewrite $pdp_{sc,2}$ as
 	$pdp_{sc,2} = P_{1}^{q_{1}} + (1-P_{1})P_{2}^{q_{2}}F_{2}$, where $F_{2} = \beta_{E}^{(1,2)}$. Similarly, the PDP expression for a $3$-hop network can be written as $pdp_{sc,3} = pdp_{sc,1h} + pdp_{sc,2h} + pdp_{sc,3h}$, where
 	\begin{eqnarray}
 	\label{PDF_semi_cumm_1_2_3_hop}
 	pdp_{sc,3h} = (1-P_{1})(1-P_{2})P_{3}^{q_{3}} \bigg(\bigg(\sum_{i=1}^{q_{1}} P_{1}^{q_{1}-i}\bigg) \sum_{i=1}^{q_{2}}P_{2}^{q_{2}-i}\nonumber\\
 	P_{3}^{i-1} +  \sum_{i=1}^{q_{1}} P_{1}^{q_{1}-i} \sum_{k=0}^{i-2}P_{2}^{q_{2}+k}  \bigg), \nonumber
 	\end{eqnarray}
 	captures the probability that the packet is dropped at the third link. In the above expression, the term $\left(\sum_{i=1}^{q_{1}} P_{1}^{q_{1}-i}\right) \left(\sum_{i=1}^{q_{2}}P_{2}^{q_{2}-i}P_{3}^{i-1}\right)$ represents the probability that the second node passes the packet without using the residual ARQs from the first node, whereas the third node makes use of all the residual ARQs of the second node. Here, $\beta_{N}^{1} \triangleq \left(\sum_{i=1}^{q_{1}} P_{1}^{q_{1}-i}\right)$ is referred to as the no borrowing term from the first node, whereas $ \beta_{E}^{2, 3} \triangleq \sum_{i=1}^{q_{2}}P_{2}^{q_{2}-i}P_{3}^{i-1}$ is the external borrowing term as defined in the two-hop case. Along the same lines, the term $\beta_{I}^{1, 2} \triangleq \sum_{i=1}^{q_{1}} P_{1}^{q_{1}-i} \sum_{k=0}^{i-2}P_{2}^{q_{2}+k}$ represents the probability that the second node passes the packet to the third node after using all the residual ARQs of the first node. We refer to this term as the internally borrowing term. In short, the PDP expression for the three-hop can be written as
 	\begin{eqnarray}
 	\label{PDF_semi_cumm_3_beta_hop}
 	pdp_{sc,3} = P_{1}^{q_{1}} + (1-P_{1})P_{2}^{q_{2}} F_{2} + (1-P_{1})(1-P_{2}) P_{3}^{q_{3}}F_{3}, \nonumber
 	\end{eqnarray} 
 	where $F_{2}=  \beta_{E}^{(1,2)}$ and $F_{3} = \beta_{N}^{(1)}\beta_{E}^{(2,3)} + 
 	\beta_{I}^{(1,2)}$. In general, the PDP expression for an $N$-hop network can be written as $pdp_{sc,N} = 
 	pdp_{sc,1h} + pdp_{sc,2h}+\ldots+pdp_{sc,Nh}$, where $pdp_{sc,Nh}$ captures the probability that the packet is dropped in the last link. This implies that the packet has survived through the first set of $N - 1$ nodes. Among the preceding $N - 1$ nodes, a node passes the packet to its next node either by using a number of attempts within its allotted ARQs without having to use the residual ARQs of the preceding node, or by using a number of attempts exceeding the ARQs allotted to it, however, by using the residual ARQs of the preceding node. We shall denote these two possible ways as state $0$ and state $1$, respectively. This implies that the packet can reach the penultimate node wherein all possible states taken by the first $N-1$ nodes comes from the space $\{0, 1\}^{N-1}$. Although the number of such sequences is at most $2^{N-1}$, not all those sequence states are valid in our case. This is because the first node in the network cannot take state $1$ because it has no preceding node to borrow the ARQs. Similarly, given that the underlying protocol is of semi-cumulative nature, node-$j$, for $j >2$, cannot forward the packet in state $1$ if node-$(j-1)$ has already forwarded the packet in state $1$. This is because node-$(j-1)$ has used more ARQs allotted to it, and therefore, node-$j$ does not have any residual ARQs to its advantage. This implies that among all possible sequences $\{0, 1\}^{N-1}$, we cannot have consecutive ones. This further implies that the total number of ways in which the packet arrives at the penultimate node is equal to the number of binary sequences of length $N - 2$ that have no consecutive ones. Henceforth, let us refer to this number as $FB(N-2)$. Let us call the set of such binary sequences as $\mathcal{FB}_{N-2}$. In order to use it to obtain the ways in which packets survive till the penultimate node, we pick $\mathbf{x} \in \mathcal{FB}_{N-2}$, and then obtain a new sequence of length $N$ as $\mathbf{x'} = [0 ~\mathbf{x} ~0]$ if the last digit of $\mathbf{x}$ is 1. Similarly, we have $\mathbf{x'}= [0 ~\mathbf{x} ~1]$ if the last digit of $\mathbf{x}$ is 0. The above changes are applicable because the first bit of $\mathbf{x'}$ has to be zero because of the first node, and moreover, if the $(N-1)$-th position is zero, that means the $N$-th node can make use of its residual ARQs. However, on the other hand, if the $(N-1)$-th position is one, then the only way the $N$-th node can drop the packet is by consuming all its ARQs. When the sequence $\mathbf{x'}$ is of the form $\mathbf{b} = [b_{1}~ b_{2} ~b_{3} \ldots b_{N}]$, we can write the corresponding probability of survival as follows. Because of no consecutive ones, let us look for sub-sequences of `01' in the sequence $\mathbf{b}$. If the pattern `01' is found in the $j$-th and $(j+1)$-th positions for $j < N-2$, then use the expression $\beta_{I}^{j, j+1}$. If the pattern `01' is found in the $(N-1)$-th and $N$-th terms, then we use the expression $\beta_{E}^{N-1, N}$. Once the above expressions are placed, the rest of the zeros in the sequence are replaced by the term $\beta_{N}^{j}$ if the zero is found in the $j$-th position. Finally, we multiply these terms to get one expression. Once a sequence is replaced by the expression, we add up all the terms to obtain $F_{N}$, which corresponds to a total of $FB(N-2)$ terms. Thus, we have $pdp_{sc,Nh} = \prod_{i = 1}^{N-1}(1-P_{i})P_{N}^{q_{N}}F_{N}$. This completes the proof.
 \end{IEEEproof}
 
 \begin{corollary}
 	\label{cor1}
 	Each term in $F_{N}$ is a product of terms of the form $\beta_{I}^{j, j+1}$, $\beta_{E}^{N-1, N}$ and $\beta_{N}^{j}$, where $\beta_{E}^{N-1, N}$ can occur at most once, $\beta_{N}^{j}$ can occur at most $N - 2$ times, and $\beta_{I}^{j, j+1}$ can occur at most $\lceil \frac{N}{2} \rceil$ times.
 \end{corollary}
 
 \begin{lemma}
 	\label{lemma1}
 	With $P = \underset{1 \leq i \leq N}{\text{max}} ~P_{i}$ and when $P < \frac{1}{2}$, we have 
 	\begin{equation}
 	\label{eq1_lemma1}
 	\beta_{E}^{N-1, N} < 2P \left( \sum_{j = 1}^{q_{N-1}} P_{N-1}^{q_{N-1} - j}\right),
 	\end{equation}
 	when $q_{N-1} > 1$. Similarly, we have 
 	\begin{equation}
 	\label{eq2_lemma1}
 	\beta_{I}^{j, j+1} < P^{2} \left( \sum_{\alpha = 1}^{q_{j}} P_{j}^{q_{j} - \alpha}\right)\left( \sum_{\gamma = 1}^{q_{j+1}} P_{j+1}^{q_{j+1} - \gamma}\right),
 	\end{equation}
 	when $q_{j+1} > 1$.
 \end{lemma}
 \begin{IEEEproof}
 	From the definition, we have $\beta_{E}^{N-1, N} = \sum_{i=1}^{q_{N-1}}P_{N-1}^{q_{N-1}-i}P_{N}^{i-1}$. Since $P = \max_{i} P_{i}$, we can upper bound it as $\beta_{E}^{N-1, N} < \sum_{i=1}^{q_{N-1}}P^{q_{N-1}-1} = q_{N-1}P^{q_{N-1}-1}$. Finally, since $q_{N-1} > 1$, we have $\beta_{E}^{N-1, N} < 2P$, and therefore \eqref{eq1_lemma1} also holds good. Similarly, from the definition, we have $\beta_{I}^{j, j+1} = \sum_{\alpha=1}^{q_{j}} P_{j}^{q_{j}-\alpha} \sum_{\gamma=0}^{\alpha-2}P_{j+1}^{q_{j+1}+\gamma}$. Using $P = \max_{i} P_{i}$, we have $\sum_{\gamma=0}^{\alpha-2}P_{j+1}^{q_{j+1}+\gamma} \leq \sum_{\gamma=0}^{\alpha-2}P^{q_{j+1}+\gamma}$. Furthermore, we have $\sum_{\gamma=0}^{\alpha-2}P_{j+1}^{\gamma} \leq \sum_{\gamma=0}^{\infty}P^{\gamma} < \frac{1}{1 - P} < 2$, wherein the last inequality holds since $P < \frac{1}{2}$. These inequalities imply that $\beta_{I}^{j, j+1} < \left(\sum_{\alpha=1}^{q_{j}} P_{j}^{q_{j}-\alpha}\right) 2P^{q_{j+1}}$. Therefore, when $q_{j+1} > 1,$ we have $\beta_{I}^{j, j+1} < \left(\sum_{\alpha=1}^{q_{j}} P_{j}^{q_{j}-\alpha}\right) 2P^{2}$, and thus \eqref{eq2_lemma1} also holds good. This completes the proof.
 \end{IEEEproof}
 

 \begin{theorem}  
 	\label{thm2}
 	For a given $\mathbf{q} = [q_{1},q_{2},\dots,q_{N}]$, at high SNR values, the PDP of the semi-cumulative scheme is upper bounded by the PDP of non-cooperative scheme. 
 \end{theorem}	
 \begin{IEEEproof}
 	We will prove this theorem by using the method of induction. For $N=2$, the PDP expression for the non-cooperative scheme is $pdp_{nc,2}= P_{1}^{q_{1}} + (1-P_{1}^{q_{1}})P_{2}^{q_{2}}$. Similarly, the PDP expression for the semi-cumulative scheme is $pdp_{sc,2} = P_{1}^{q_{1}} + (1-P_{1})\bigg(\sum_{i=1}^{q_{1}}P_{1}^{q_{1}-i}P_{2}^{q_{2}+i-1}\bigg).$ When $q_{1} > 1$, note that $P_{2}^{i-1} < 1$ for $i = 2, 3, \ldots, q_{1}$, and therefore, we have $pdp_{sc,2} < pdp_{nc,2}$. This completes the proof for $N = 2$. Assuming that the statement of the theorem is true for any $k$-hop network, we will prove the result for a $(k+1)$-network. The PDP expression of the semi-cumulative scheme for $(k+1)$-hop network is $pdp_{sc,k+1}= pdp_{sc,k} + F_{k+1}P_{k+1}^{q_{k+1}}\prod_{i=1}^{k}(1-P_{i}),$ where $pdp_{sc,k}$ is the PDP of the semi-cumulative scheme for the $k$-hop network. From induction, we have $pdp_{sc,k} < pdp_{nc,k}$, and therefore, we only need to prove that $F_{k+1}\prod_{i=1}^{k}(1-P_{i}) < \prod_{i=1}^{k}(1-P_{i}^{q_{i}})$. In other words, we need to prove that $F_{k+1} < \prod_{i=1}^{k}\left(\sum_{j = 1}^{q_{i}}P_{i}^{q_{i} - j}\right)$. From Theorem \ref{thm1}, we know that $F_{k+1}$ can be written using binary sequences of length $k+1$ such that each sequence does not contain consecutive ones. Furthermore, we have shown that each term of $F_{k+1}$ is a product of several terms of the form $\beta_{I}^{j, j+1}$, $\beta_{N}^{j}$, for $j < k - 1$, and $\beta_{E}^{k, k+1}$. From Corollary \ref{cor1}, it is clear that there is only one term in $F_{k+1}$ which has $\beta_{E}^{k, k+1}$ appearing once in conjunction with $\prod_{\alpha = 1}^{k -1}\beta_{N}^{\alpha}$, and all other terms either have both $\beta_{E}^{k, k+1}$ and $\beta_{I}^{j, j+1}$, or only $\beta_{I}^{j, j+1}$. In addition, from Lemma \ref{lemma1}, this implies that $F_{k+1}$ can be upper bounded as $F_{k+1} < \prod_{i=1}^{k}(1-P_{i}^{q_{i}})\eta(P)$ where $\eta(P)$ is a polynomial in $P$ of the form $2P + 2P^{2}(k+1-2) +\rho(P)$ such that $\rho(P)$ is a polynomial in $P$ of degree at least three. At high SNR values, it is clear that $P <<1$, and therefore, we can show that $\eta(P) < 1$. This, in turn, implies that $F_{k+1} < \prod_{i=1}^{k}(1-P_{i}^{q_{i}}).$ This completes the proof.
 \end{IEEEproof}	

Given that the expression for $pdp_{sc, N}$ is obtained, in the subsequent sections, we propose low-complexity algorithms to solve Problem \ref{opt_problem} since implementing exhaustive search to find the optimal distribution of ARQs it is not practically feasible.
 
 \section{Optimal ARQ Distribution of the Semi-Cumulative Scheme}
 \label{sec:opt_distribution}
 
For an $N$-hop network with $\mathbf{q}= [q_{1},q_{2},\ldots,q_{N-1}, q_{N}]$, suppose that the ARQs for the first $N-2$ hops are fixed, and we are interested in computing the optimal values of $q_{N-1}$ and $q_{N}$ that minimizes the PDP. If we start with $\tilde{\mathbf{q}}= [q_{1},q_{2},\ldots,0, q_{N-1}+q_{N}]$, it may give us a sub-optimal PDP. Therefore, using $\tilde{\mathbf{q}}$, as we keep transferring one ARQ from the last node to the penultimate node, we can expect the PDP to decrease, and then start to increase beyond a certain number of transfers. Towards understanding this transition of PDP, we are interested in understanding the structure of the ARQ distribution when the PDPs of network with $\mathbf{q}= [q_{1},q_{2},\ldots,q_{N-1}, q_{N}]$ and $\mathbf{q}^{'}= [q_{1},q_{2},\ldots,q_{N-1}+1, q_{N}-1]$ are equal. Once we obtain this relation, we can analytically compute the values of $q_{N-1}$ and $q_{N}$ for a given $q_{1}, q_{2}, \ldots, q_{N-2}$, which in turn reduces the search space for computing the optimal ARQ distribution. This result is formally captured in the following theorem.

 \begin{theorem}
 \label{thm3}
 To find the optimal distribution of ARQs for a given $N$-hop network, brute force search for an $N$-hop network can be reduced into brute force search for $(N-2)$-hop network by fixing ARQs $q_{1}, q_{2}, \ldots, q_{N-2}$. 
 \end{theorem}	
 \begin{IEEEproof}
 	Consider an $N$-hop semi-cumulative scheme with $\mathbf{q}= [q_{1},q_{2},\ldots, q_{N}]$ where $\sum_{i=1}^{N}q_{i}= q_{sum}$. Let $pdp_{sc, N}$ and $pdp_{sc,N}^{'}$ represent the PDP of the $N$-hop network with $\mathbf{q}= [q_{1},q_{2},\ldots,q_{N-1}, q_{N}]$ and $\mathbf{q}^{'}= [q_{1},q_{2},\ldots,q_{N-1}+1, q_{N}-1]$ respectively. The PDP expressions with $\mathbf{q}$ and $\mathbf{q}^{'}$ can be respectively written as
 	\begin{eqnarray}
 	pdp_{sc,N} &= & pdp_{sc,1h} + \ldots + pdp_{sc,(N-1)h} + pdp_{sc,Nh}.\nonumber \\
 	& = & pdp_{sc,N-2} + pdp_{sc,(N-1)h}+ pdp_{sc,Nh},\nonumber \\
 	pdp_{sc,N}^{'} & = & pdp_{sc,1h}^{'} + \dots + pdp_{sc,(N-1)h}^{'}+ pdp_{sc,Nh}^{'}.\nonumber\\
 	&= &pdp_{sc,N-2}^{'} + pdp_{sc,(N-1)h}^{'}+ pdp_{sc,Nh}^{'},\nonumber  
 	\end{eqnarray}
 	where the individual expressions are the probabilities that the packet is dropped at the intermediate links. It is straightforward to note that $pdp_{sc,jh} = pdp_{sc,jh}^{'}$ for $1\leq j \leq N-2$ since the first $N-2$ terms are the same in $\mathbf{q}$ and $\mathbf{q}^{'}$. Therefore, on equating $pdp_{sc,N}= pdp_{sc,N}^{'}$, we get 
 	\begin{eqnarray}
 	pdp_{sc,(N-1)h}+ pdp_{sc,Nh}= pdp_{sc,(N-1)h}^{'}+ pdp_{sc,Nh}^{'},\nonumber\\
 	pdp_{sc,(N-1)h}- pdp_{sc,(N-1)h}^{'} =-(pdp_{sc,Nh} - pdp_{sc,Nh}^{'}),\nonumber  
 	\end{eqnarray}
 	where we can write $pdp_{sc,(N-1)h}^{'}= P_{N-1}\left(pdp_{sc,(N-1)h}\right)$ because at the $(N-1)$-th hop, every term of $F_{N-1}^{'}$ gets multiplied by $P_{N-1}$ since one ARQ has been transferred from the $N$-th hop.  Hence, we can write
 	\begin{eqnarray}
 	pdp_{sc,(N-1)h}(1-P_{N-1}) &=&-(pdp_{sc,Nh} - pdp_{sc,Nh}^{'}).\nonumber  
 	\end{eqnarray} 
 	On expanding the above equation and including $(1-P_{N-1})$ in the product loop, we write
 	\begin{eqnarray}
 	& &\bigg(\prod_{i=1}^{N-1}(1-P_{i})P_{i}^{q_{i}}\bigg)\bigg(\frac{F_{N-1}}{\prod_{i=1}^{N-2}P_{i}^{q_{i}}} \bigg)  = \bigg(\prod_{i=1}^{N-1}(1-P_{i})\nonumber\\
 	& &P_{i}^{q_{i}}\bigg)P_{N}^{q_{N}}\frac{(P_{N}^{-1}F_{N}^{'}-F_{N})}{\prod_{i=1}^{N-1}P_{i}^{q_{i}}} , \nonumber
 	\end{eqnarray}
 	where $F_{N}^{'}$ is the term obtained using the Fibonacci series corresponding to $pdp_{sc,Nh}^{'}$. The above equality can be further simplified as
 	\begin{eqnarray}
 	\label{eq:indept}
 	& & \bigg(\frac{F_{N-1}}{\prod_{i=1}^{N-2}P_{i}^{q_{i}}} \bigg)  = P_{N}^{q_{N}}\frac{(P_{N}^{-1}F_{N}^{'}-F_{N})}{\prod_{i=1}^{N-1}P_{i}^{q_{i}}}.
 	\end{eqnarray}
 	In the rest of the proof, we will show that $\frac{P_{N}^{-1}F_{N}^{'}-F_{N}}{P_{N-1}^{q_{N-1}}}$ does not contain $q_{N-1}$ in it. Towards that direction, note that both $F_{N}^{'}$ and $F_{N}$ contain the same number of terms in their expansion using Fibonacci series, however, with the difference that the terms $q_{N}$ and $q_{N-1}$ in $F_{N}$ appear as $q_{N} - 1$ and $q_{N-1} + 1$ in $F_{N}^{'}$, respectively. When constructing $F_{N}^{'}$ and $F_{N}$ using binary sequences of length $N$, we partition the terms of $F_{N}^{'}$ and $F_{N}$ into two categories, namely: the sequences that end with `01' and sequences that end with `10'. This is because the states of the nodes before the last two digits are the same for both $F_{N}^{'}$ and $F_{N}$. As a result, for the sequences that end with `01', we can take the term $\beta_{E}^{N-1, N}$ common, and only focus on  its effect in $\frac{F_{N}^{'}-P_{N}F_{N}}{P_{N-1}^{q_{N-1}}}$. Similarly, for the sequences that end with `10', we can take the term $\beta_{I}^{N-2, N-1}$ common, and only focus on its effect in $\frac{P_{N}^{-1}F_{N}^{'}-F_{N}}{P_{N-1}^{q_{N-1}}}$. To handle the former case, the term $\beta_{E}^{N-1, N}$ from $F_{N}^{'}$ is of the form $\sum_{i=1}^{q_{N-1} + 1}P_{N-1}^{q_{N-1}+1-i}P_{N}^{i-1} = P_{N-1}^{q_{N-1}} \left(\sum_{i=1}^{q_{N-1} + 1}P_{N-1}^{1-i}P_{N}^{i-1}\right)$, whereas the term $\beta_{E}^{N-1, N}$ from $F_{N}$ is of the form $\sum_{i=1}^{q_{N-1}}P_{N-1}^{q_{N-1}-i}P_{N}^{i-1} = P_{N-1}^{q_{N-1}} \left(\sum_{i=1}^{q_{N-1}}P_{N-1}^{-i}P_{N}^{i-1}\right)$. Therefore, the difference of the two corresponding terms in $\frac{P_{N}^{-1}F_{N}^{'}-F_{N}}{P_{N-1}^{q_{N-1}}}$ is $\frac{1}{P_{N}}$, and this is because of the equality 
 	\begin{eqnarray}
 	\label{PDF_semi_cumm_3_formulae_1}
 	\sum_{i=1}^{q_{N-1}}P_{N-1}^{-i}P_{N}^{i-1} - \sum_{i=1}^{q_{N-1}+1}P_{N-1}^{1-i}P_{N}^{i-2} = - \frac{1}{P_{N}}.
 	\end{eqnarray}
 	This completes the proof that $\frac{P_{N}^{-1}F_{N}^{'}-F_{N}}{P_{N-1}^{q_{N-1}}}$ does not contain $q_{N-1}$ in it from sequences ending with `01'. To handle the sequences that end with `10', the term $\beta_{I}^{N-2, N-1}$ contributing to $F_{N}^{'}$ is of the form $\sum_{i=1}^{q_{N-2}} P_{N-2}^{q_{N-2}-i} \sum_{k=0}^{i-2}P_{N-1}^{q_{N-1}+ 1 + k}$. Similarly, the term $\beta_{I}^{N-2, N-1}$ contributing to $F_{N}$ is of the form $\sum_{i=1}^{q_{N-2}} P_{N-2}^{q_{N-2}-i} \sum_{k=0}^{i-2}P_{N-1}^{q_{N-1} + k}$. Therefore, when evaluated as $P_{N}^{-1}F_{N}^{'}-F_{N}$, the term  $P_{N-1}^{q_{N-1}}$ can be taken common from both the terms, and therefore, the term $\frac{F_{N}^{'}-P_{N}F_{N}}{P_{N-1}^{q_{N-1}}}$ does not contain $q_{N-1}$ in it from sequences ending with `10'.
 	 	
 	Henceforth, \eqref{eq:indept} is written as $R_{1,N} = P_{N}^{q_{N} }R_{2, N}$, wherein $R_{1, N} \triangleq \bigg(\frac{F_{N-1}}{\prod_{i=1}^{N-2}P_{i}^{q_{i}}} \bigg)$ and $R_{2, N} \triangleq  \frac{(P_{N}^{-1}F_{N}^{'}-F_{N})}{\prod_{i=1}^{N-1}P_{i}^{q_{i}}}$ do not contain the terms $P_{N}^{q_{N}}$ and $q_{N-1}$. Hence, $R_{1,N}$ and $R_{2,N}$ are constants since $\{P_{i} ~|~i = 1,2,\dots,N\}$ and $\{q_{i} ~|~ i = 1,2,\dots,N-2\}$ are fixed. Now, we can rewrite the equality condition as $P_{N}^{q_{N}}= \frac{R_{1,N}}{R_{2,N}}$, or as $q_{N} = \frac{\big(\log \frac{R_{1,N}}{R_{2,N}}\big)}{\log P_{N}}$. Note that in our work, we have a condition that $q_{i} \in \mathbb{Z}_{+} $, however, the solution of $q_{N}= \frac{\big(\log \frac{R_{1,N}}{R_{2,N}}\big)}{\log P_{N}}$ may belong to $\mathbb{R}$. It implies that to find the optimal solution which lies in $\mathbb{Z}_{+}$, we need to obtain either $\ceil{q_{N}}$ or $\floor{q_{N}}$ from the equality condition. It can be observed that $\ceil{q_{N}}$ will decrease $P_{N}^{\ceil{q_{N}}}$, and this implies that $pdp_{sc,N} > pdp_{sc,N}^{'}$, and this is a sub-optimal solution because when we give one more ARQ from the last hop to the second last hop, PDP decreases. On the other hand, if we use $\floor{q_{N}}$, then $P_{N}^{\floor{q_{N}}}$ increases, which implies $pdp_{sc,N} < pdp_{sc,N}^{'}$. Therefore, on giving one more ARQ from the last hop to second last hop, PDP increases, and this implies that using $q_{N}= \lfloor \frac{\big(\log \frac{R_{1,N}}{R_{2,N}}\big)}{\log P_{N}} \rfloor$ in $\mathbf{q}$ captures the optimal solution conditioned on the first $N-2$ ARQ numbers. Thus, on fixing $q_{1},q_{2},\dots,q_{N-2}$, we can analytically compute $q_{N}$, and also compute $q_{N-1}$ using the relation $q_{N-1} = q_{sum} - \sum_{t = 1, t \neq N-1}^{N} q_{t}$.
 \end{IEEEproof}


%

\section{Low-complexity Algorithms}
\label{sec:proposed_method}
From Theorem \ref{thm3}, we have proved that the search space for the $N$-hop network can be reduced to the search space of an $(N-2)$-hop network. Henceforth, we refer to this reduction as a one-fold technique. For a large value of $N$, we observe that the one-fold technique may not be feasible to implement in practice. Therefore, we propose low-complexity algorithms to further reduce the search space for the optimization problem.

\subsection{List Generation using Multi-folding}
\label{sec:multi-algo}

In the proposed multi-folding algorithm, as presented in Algorithm \ref{multi-level-algo}, instead of folding the network once from $N$-hop to $(N-2)$-hop, we fold it multiple times to $(N-4)$-hop, $(N-6)$-hop and so on up to a $2$-hop network or a $1$-hop network depending on whether $N$ is even or odd, respectively. When the network is reduced (or folded) to a $j$-hop network, we need to provide a sum of $\tilde{q}_{sum, j}$ ARQs to it, and it is clear that $\tilde{q}_{sum,j}$ can take all possible values in a range $[j, q_{sum}-(N-j)+1]$. When folding the network up to $j$-hops, for $j \geq 4$ and $j \geq 3$ when $N$ is even and odd, respectively, we fix the ARQs for the first $(j-2)$-hops and then compute $q_{j - 1}$ and $q_{j}$ using Theorem \ref{thm3} for each value of $\tilde{q}_{sum,j}$. Subsequently, we create a list of ARQ distributions [$q_{1},\ldots,q_{j}$], denoted by $\mathcal{L}_{j}$, by varying the values of $\tilde{q}_{sum,j}$. Following a similar procedure, the candidates of $\mathcal{L}_{j}$ are used to generate $\mathcal{L}_{j + 2}$ for the $(j + 2)$-hop network by using Theorem \ref{thm3} for each value of $\tilde{q}_{sum,j + 2}$. This way, a list of ARQ distributions are obtained through $\mathcal{L}_{N}$ for the original $N$-hop network. It is clear that the size of the search space $\mathcal{L}_{N}$ reduces with increase in the number of folds.

\subsection{Multi-folding based Greedy Algorithm}
\label{sec:greedy-algo}

To further reduce the size of the search space from that of Algorithm \ref{multi-level-algo}, we propose to retain the ARQ distribution that gives us minimum PDP for a given $\tilde{q}_{sum,j}$ from the list $\mathcal{L}_{j}$. This way, only one ARQ distribution survives for a given $\tilde{q}_{sum,j}$, thereby significantly reducing the list size when the algorithm traverses to $\tilde{q}_{sum,N}$. In the process of obtaining $q_{j}$ for each $\tilde{q}_{sum,j}$, we note that only the floor of the ratio $\frac{\log R_{j}}{\log P_{j}}$ is chosen to obtain $q_{j}$. However, by observing that the optimal distribution of the folded network may not contribute to the optimal distribution of the original $N$-hop network, we also propose to select the ARQ distribution by ceiling the ratio $\frac{\log R_{j}}{\log P_{j}}$, where $R_{j}= \frac{R_{1,j}}{R_{2,j}}$. In other words, for each $\tilde{q}_{sum,j}$ in $\mathcal{L}_{j}$ we choose the ARQ distribution that minimizes the PDP for the $j$-hop network, and for that selected ARQ distribution, we also pick the ARQ distribution obtained by giving one ARQ from the last node to the penultimate node. It is straightforward to observe that this technique gives us a significantly shorter list compared to the multi-folding approach.  
\begin{algorithm}
\caption{\label{multi-level list agorithm}Multi-folding list algorithm}
\label{multi-level-algo} 
\begin{algorithmic}[1]
\Require  $N$, $q_{sum}$, $\mathbf{P} = [P_{1}, P_{2}, \ldots, P_{N}]$.
\Ensure $\mathcal{L}_{final} $ - List of ARQ distributions in search space.
\State $\mathcal{L}_{k} = \{\phi\}$ (A null set)  $\ \forall \ k= 1,2,\ldots,N$. 
\If {$N= odd$} 
\State Start with fixing $q_{1}$.
\State $\mathcal{L}_{1} = \{ [1, q_{sum}-(N-1)+1 ]\}$.
\State Assign $p=3$.
\For {$j=p:2:N$}
\State Assign $C=1$.
\For {$i_{1}=1:|\mathcal{L}_{j-2}|$}
\State $[q_{1}, \ldots, q_{j-2}] = \mathcal{L}_{j-2}({i_{1}})$
\State Compute $q_{j} = \floor{\frac{\log R_{j}}{\log P_{j}}}$ where $R_{j} = \frac{R_{1,j}}{R_{2,j}}$.
\For {$\tilde{q}_{sum,j} = j: (q_{sum}-(N-j)+1)$}.
\State Compute $q_{j-1} = \tilde{q}_{sum,j} - \sum_{t=1, t\neq j-1}^{j}q_{t}$.
\If{$q_{j-1} \geq 0$}
\State Insert [$\mathcal{L}_{j-2}(i_{1})||q_{j-1}||q_{j}$] into $\mathcal{L}_{j}(C)$.
\State Assign $C=C+1$.
\EndIf
\EndFor
\EndFor
\EndFor
\State $\mathcal{L}_{final}= \{\mathcal{L}_{N} | q_{j+1} \neq 0\ \text{for} \ q_{j} = 0 \ \text{or} \ 1  \} $.
\ElsIf{$N = even$}
\State Start with fixing $q_{1}$ and $q_{2}$.
\State $\mathcal{L}_{2} = \{\{q_{1},q_{2}\} \in \mathbb{Z}_{+}^{2} | q_{1}+q_{2} \in [2, q_{sum}-(N-2)+1 ]\}$
\State Assign $p=4$.
\State Repeat steps from line number $6$ to $20$.
\EndIf
\end{algorithmic}
\end{algorithm}



\begin{figure}
\centering  \includegraphics[scale = 0.42]{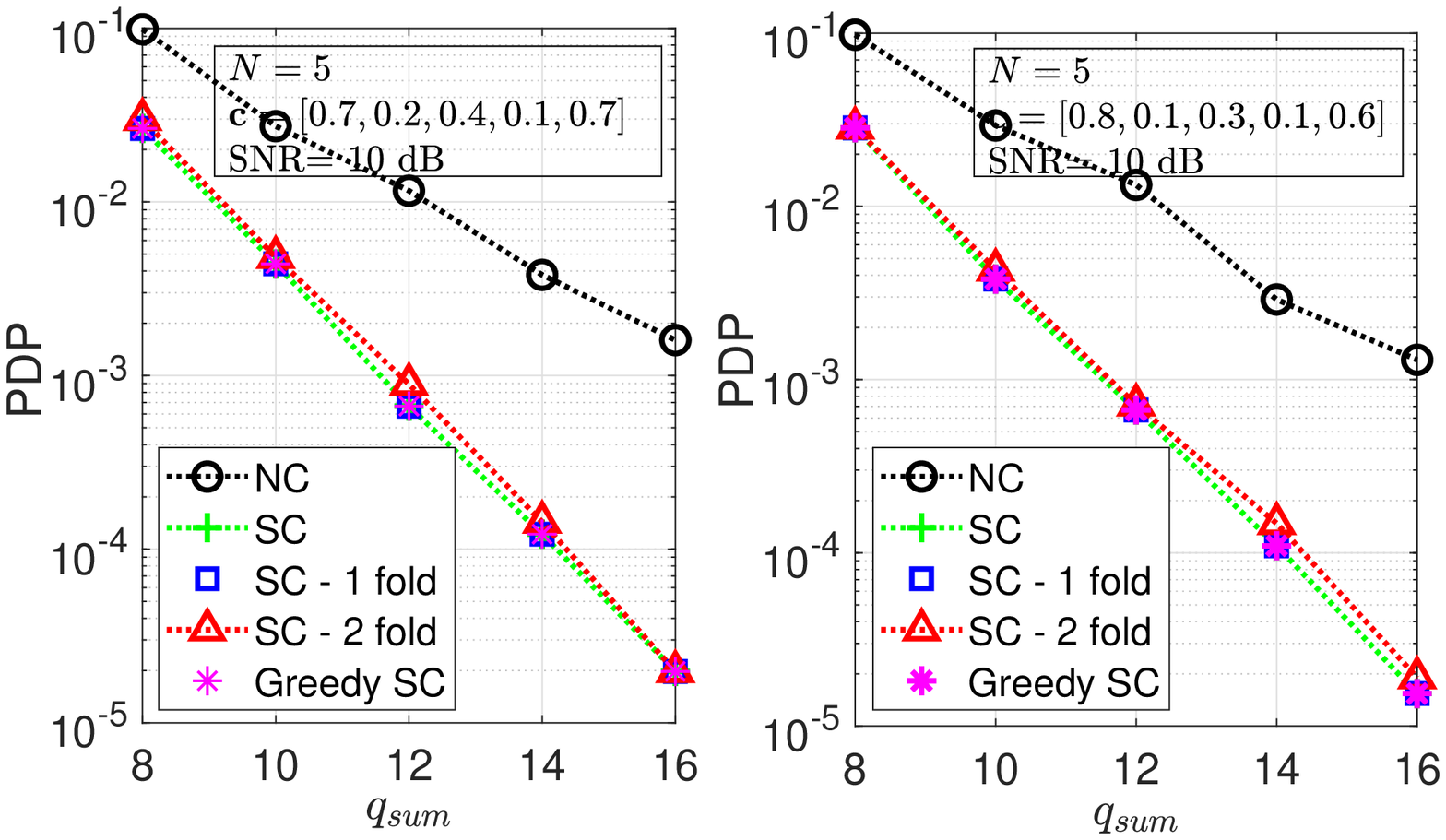}
\vspace{-0.4cm}
\centering{\caption{PDP comparison for our strategies for $N = 5$ at $R = 1$.  
\label{PDP_N_5}}}
\centering \includegraphics[scale = 0.422]{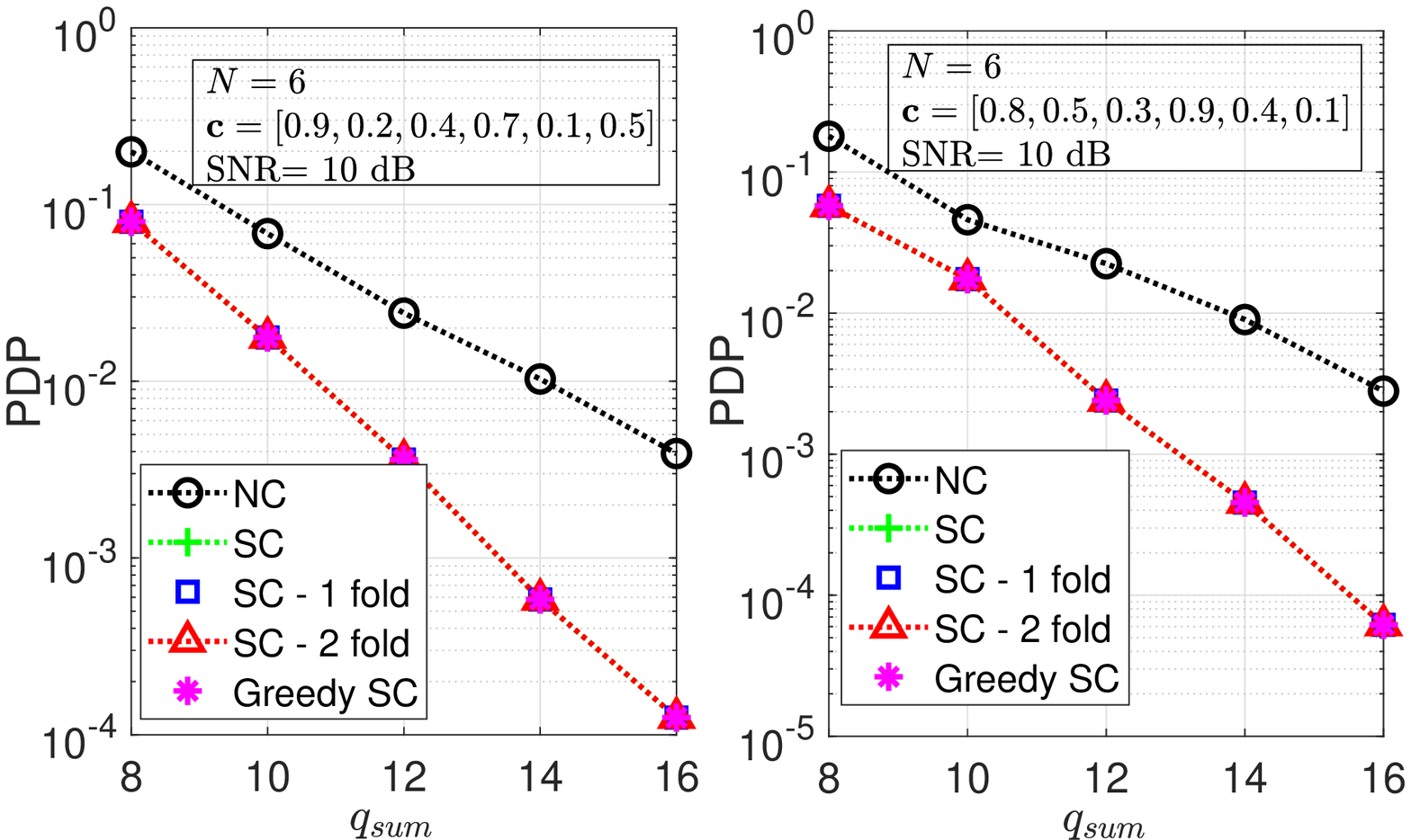}\centering{\caption{PDP comparison for $N = 6$ at $R = 1$.   
\label{PDP_N_6}}}
\end{figure}	

\begin{figure}
\centering \includegraphics[scale = 0.42]{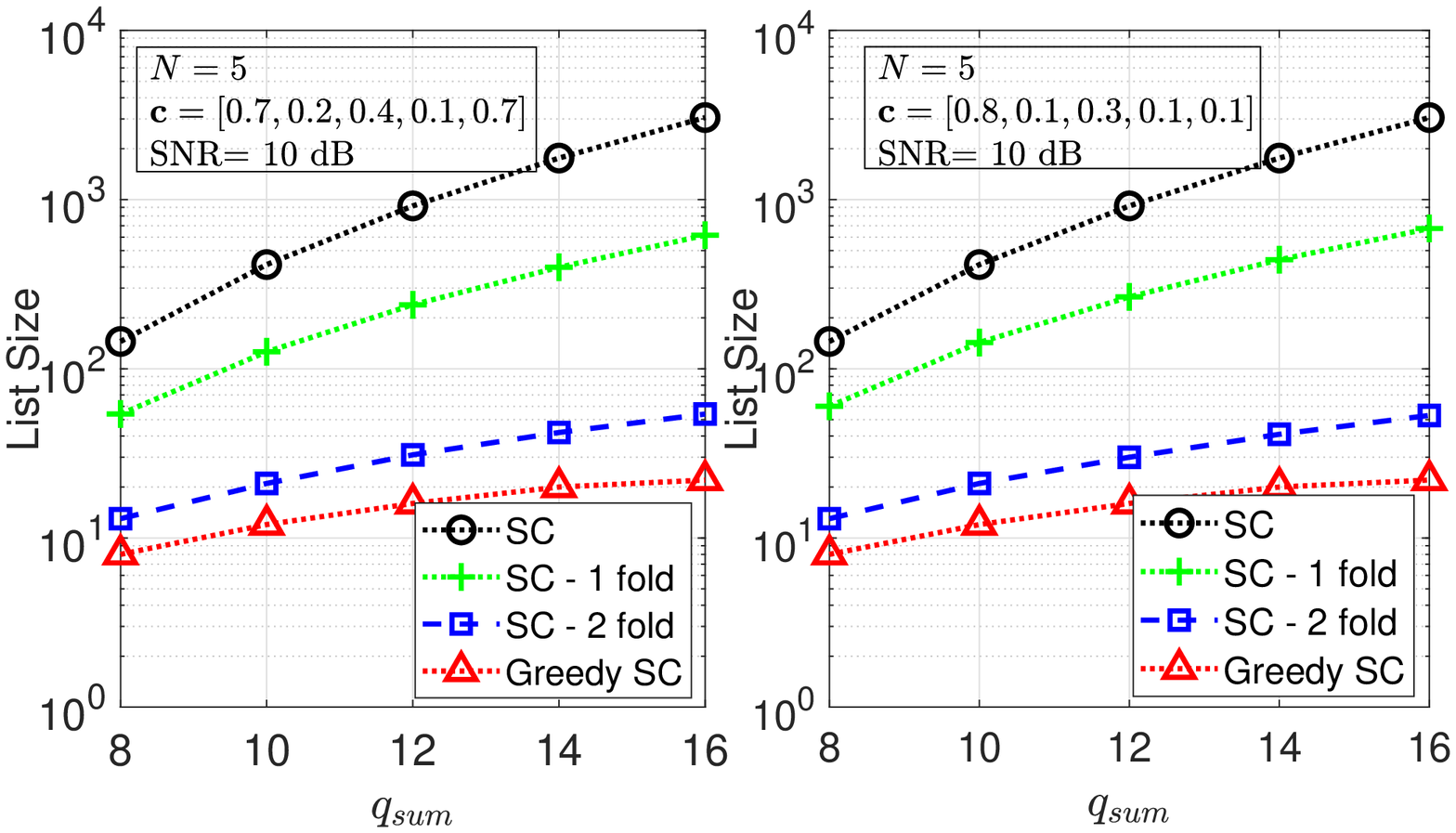}
\centering{\caption{Comparison of list sizes for $N = 5$ at $R = 1$. 
\label{Complexity_5H}}}
\vspace{0.19cm}
\centering  \includegraphics[scale = 0.42]{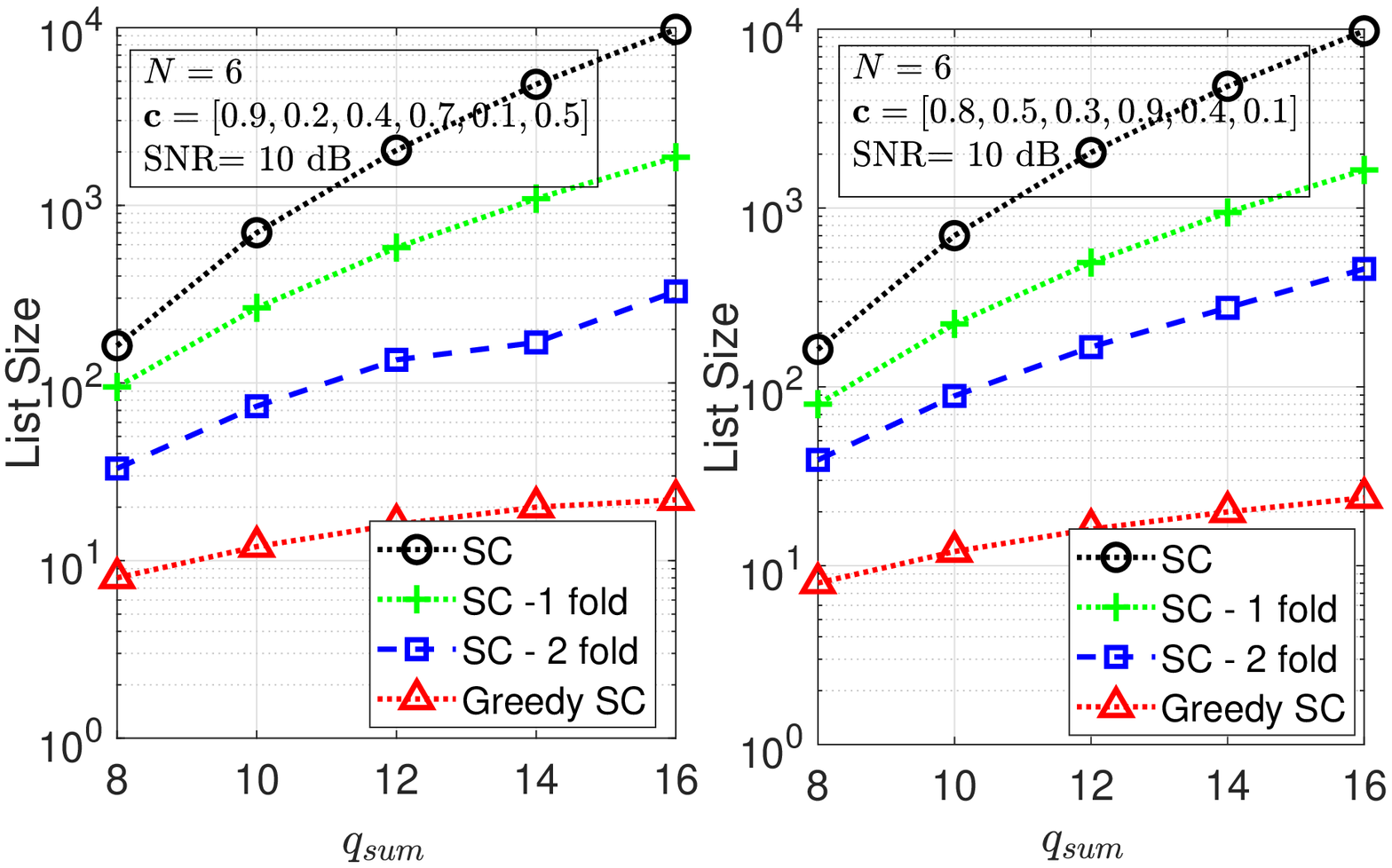}
\centering{\caption{List size for $N = 6$ at $R = 1$.  
\label{Complexity_6H}}}
\end{figure}


\section{Simulation Results and Complexity Analysis}
\label{sec:Sims}

In this section, we present simulation results to analyse the PDP of the semi-cumulative scheme for various values of $N, q_{sum}$, and LOS vectors. First, in Fig. \ref{PDP_N_5} and Fig. \ref{PDP_N_6}, we present simulation results to compare the PDP of the semi-cumulative scheme with that of the non-cooperative scheme \cite{our_work_1}. Although we have proved the dominance of our strategy theoretically, the plots confirm  that the PDP of the semi-cumulative scheme outperforms the PDP of the non-cooperative scheme with no or negligible increase in the overhead. Furthermore, to showcase the benefits of using the multi-fold algorithm and the greedy algorithm, we plot the PDP offered by these algorithms for $N=5$ and $N=6$ in Fig. \ref{PDP_N_5} and Fig. \ref{PDP_N_6}, respectively. The plots confirm that while the multi-fold algorithm provides near-optimal ARQ distribution, the greedy algorithm is successful in offering the optimal ARQ distributions for our simulation parameters.     

In terms of complexity, for an $N$-hop network, the size of the search space for the semi-cumulative scheme is upper bounded by $\binom{q_{sum}+N-1}{N-1}$. However, with the multi-fold algorithm, we have shown that the search space can be reduced. To showcase the reduction, we plot the size of the search space ($\mathcal{L}_{N}$) of the multi-fold algorithm for $N=5$ and $N=6$. For these cases, since we can fold the network at most twice, we have shown the results for both one-fold and two-fold cases. The simulation results, as shown in Fig. \ref{Complexity_5H} and Fig. \ref{Complexity_6H}, display significant reduction in the list size as we move to one-fold and two-fold. Finally, the plots also show that the list size of the greedy algorithm is shorter than the multi-fold case, and it is, therefore, amenable to implementation in practice. 
  
\section{Summary}
\label{sec:conclusions}
In this work, we have proposed a semi-cumulative ARQ scheme wherein intermediate relays of the network can use the residual ARQs from their previous node thereby reducing the PDP of the network with no compromise in the latency constraints. We have derived closed-form expressions on the PDP of the semi-cumulative scheme for any given value of $N$ and $q_{sum}$, and have subsequently addressed solving the optimization problem of minimizing the PDP under a sum constraint on the total number of ARQs. 

\section*{Acknowledgements}
This work was supported by the Indigenous 5G Test Bed project from the Department of Telecommunications, Ministry of Communications, New Delhi, India.

\end{document}